\documentclass{emulateapj}

\usepackage{natbib}
\slugcomment{}

\shorttitle{NGC~4449:CL~77}
\shortauthors{Annibali et al.}

\begin{document}


\title{Cluster 77 in NGC~4449: The nucleus of a satellite galaxy being transformed  into a globular cluster? \footnote{Based on observations 
with the NASA/ESA {\it Hubble Space Telescope}    obtained at the 
Space Telescope Science Institute    which is operated  by AURA    Inc.   
for NASA under contract NAS5-26555.}}



\author{F. Annibali\altaffilmark{2}, M. Tosi\altaffilmark{2}, A. Aloisi\altaffilmark{3}, R.~P. van der Marel\altaffilmark{3}, D. Martinez-Delgado\altaffilmark{4} } 

\altaffiltext{2}{INAF-Osservatorio Astronomico di Bologna, Via Ranzani 1, I-40127 Bologna, Italy}

\altaffiltext{3}{Space Telescope Science Institute, 3700 San Martin Drive, Baltimore MD 21218, USA}

\altaffiltext{4}{Max-Planck-Institut f\"{u}r Astronomie, Konigstuhl, 17 D-69117 Heidelberg, Germany }


\begin{abstract}

We report the discovery in our HST ACS B, V, and I images of NGC~4449 of a globular cluster (GC) which appears 
associated with two tails of blue stars. The cluster is massive 
($M\sim1.7\times10^6 M_{\odot}$) and highly flattened ($\epsilon\sim0.24$).
 From the color-magnitude diagrams of the resolved stars we infer active star formation in the tails over the past $\sim$200 Myr. 
In a diagram of mean projected mass density inside $r_e$ versus total mass the cluster lies at the upper end of the GC distribution,  
where galaxy nuclei are. The north-west tail is associated with a concentration of HI and infrared (dust/PAHs) emission which appears as part of a much 
longer stream wrapping around the galaxy.  These properties suggest that the cluster may be the 
nucleus of a former gas-rich satellite galaxy 
undergoing tidal disruption by NGC~4449.
If so, the cluster is seen in an earlier phase compared to 
other suggested nuclei of disrupted galaxies such as $\omega$Cen (Milky Way) and G1 (M31). 

\end{abstract}





\section{Introduction}

Within the hierarchical framework for galaxy formation, accretion and disruption of dwarf galaxies 
drive the formation of larger galaxy halos \citep[e.g.][]{helmi99}. This process leaves long-lived 
relics in the form of streams of stellar remnants that remain aligned to the orbital 
path of the parent satellite \citep[e.g.][]{john96}. 
The closest examples of such processes are the tidal stream of the Sagittarius dwarf spheroidal 
galaxy (Sgr dSph) \citep{ibata94}, which is  merging with the Milky Way (MW) \citep[e.g.][]{ibata01a,maj03},  
and the Andromeda (M31) streams \citep[e.g.][]{ibata01b}.  
M54, the second most massive Galactic globular cluster (GC), lies at the photometric center 
and possibly at the distance of the Sgr dSph, and  was considered as its nucleus, although there are suggestions  that it may  actually be $\sim$2 kpc in the foreground of the galaxy center
\citep{bellazzini08,siegel11}. 

Stellar streams provide unquestionable evidence of satellite accretion, but 
there may be less manifest cases in which the former satellite galaxy is no longer identifiable and what remains after the disruption 
is just a naked nucleus. Two such potential cases are $\omega$Cen and G1, the most massive GCs in the MW and M31, respectively, that have been suggested to be nuclei of disrupted galaxies on the basis of the following properties:  
the presence of multiple stellar populations \citep[e.g.][]{pancino00,bellini10} or of a chemical abundance spread \citep{jp10,meylan01};  
 their very flattened shape \citep{pancino00,meylan01}; velocity dispersions larger than in other GCs \citep{djo97,sollima09}; 
the dynamical evidence, in G1, for the presence of a central black hole \citep{gebhardt02}; the very bound retrograde orbit of  $\omega$Cen with respect to Galactic rotation \citep{dinescu99}; a tidal plume of stars very near the (projected) position of G1 \citep{ferguson02}. 
Because of the presence of tidal tails, the GC Palomar 14 has also been suggested to be the remnant of an ancient galaxy progressively disrupted by the interaction with the  MW \citep{sollima11}.
 Both $\omega$Cen and G1 lie, in a mean projected mass density 
versus total-mass diagram,  close to the locus  of galaxy nuclei  \citep[see Fig.~3 in][]{walcher05}. 
Galaxy nuclei have very similar properties to GCs \citep[see e.g.][]{vdm07}, but are more massive than typical GCs, and exhibit mixed-age populations, generally with an underlying old stellar population. 
  
In this letter we report the discovery in the starburst irregular galaxy NGC~4449 ($D=3.82\pm0.18$ Mpc) of an old cluster surrounded by two structures of young stars roughly symmetrically located, reminiscent of tidal tails. The cluster ($\alpha_{J2000}=12 \  27 \  57.53$, $\delta_{J2000}=44 \  05 \  28.16$) is located west at a projected distance of $\sim$3 kpc from the galaxy center, has a mass of $M\sim1.7\times10^6 M_{\odot}$, a luminosity-weighted age of $\sim$7 Gyr, and exhibits significant ellipticity ($\epsilon \sim0.24$)
\citep[][hereafter A11]{ann11}. These properties suggests that it may be a galaxy nucleus still surrounded by remains of its former parent galaxy. 

The star properties  in the cluster and  in the tails are described in Section~2 through the HST/ACS color-magnitude diagrams (CMDs). Archival data of NGC~4449 at other wavelengths (ultraviolet, infrared, radio) are examined in Section~3. We discuss the arguments in favor and against CL77 being a nucleus  and summarize our results in Section~4.

\section{HST data and  color-magnitude diagrams}

\begin{center}
\begin{figure}
\epsscale{1.2}
\plotone{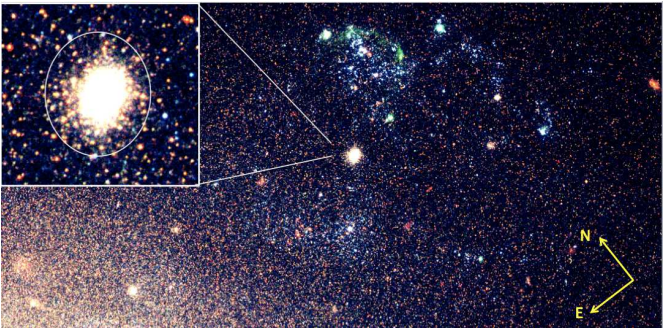}
\caption{Color-combined image (blue$=$B; green$=$V; red$=$I ) showing tails of blue stars around CL~77.  
The total FOV is $\sim$(120$\times$60) arcsec$^2$.  The (7$\times$7) arcsec$^2$ insertion  (top left)  shows resolved stars around the cluster. The ellipse corresponds to region C in Fig.~\ref{sat_cmd}. The center of the galaxy is located $\sim$80 arcsec from the bottom left of the figure. 
 \label{sat_img}}
\end{figure}
\end{center}

NGC~4449 was observed with the Wide Field Channel of the Advanced Camera for Surveys (ACS) in the F435W (B), F555W (V), F814W (I) and F658N (H$\alpha$) filters 
(GO program 10585) with a total field of view (FOV) of $\sim$380$\times$200 arcsec$^2$.
Observations and data reduction were described by \cite{ann08} (hereafter A08) and A11, where we presented respectively the CMDs of the resolved stars and a study of the cluster population in 
NGC~4449. The cluster  properties (size, ellipticity, total magnitudes, ages and 
masses) are listed in Table~1 of A11. The cluster we discuss here was labelled as 
cluster~77 in A11 (hereafter CL~77).

\begin{center}
\begin{figure*}
\epsscale{1}
\plotone{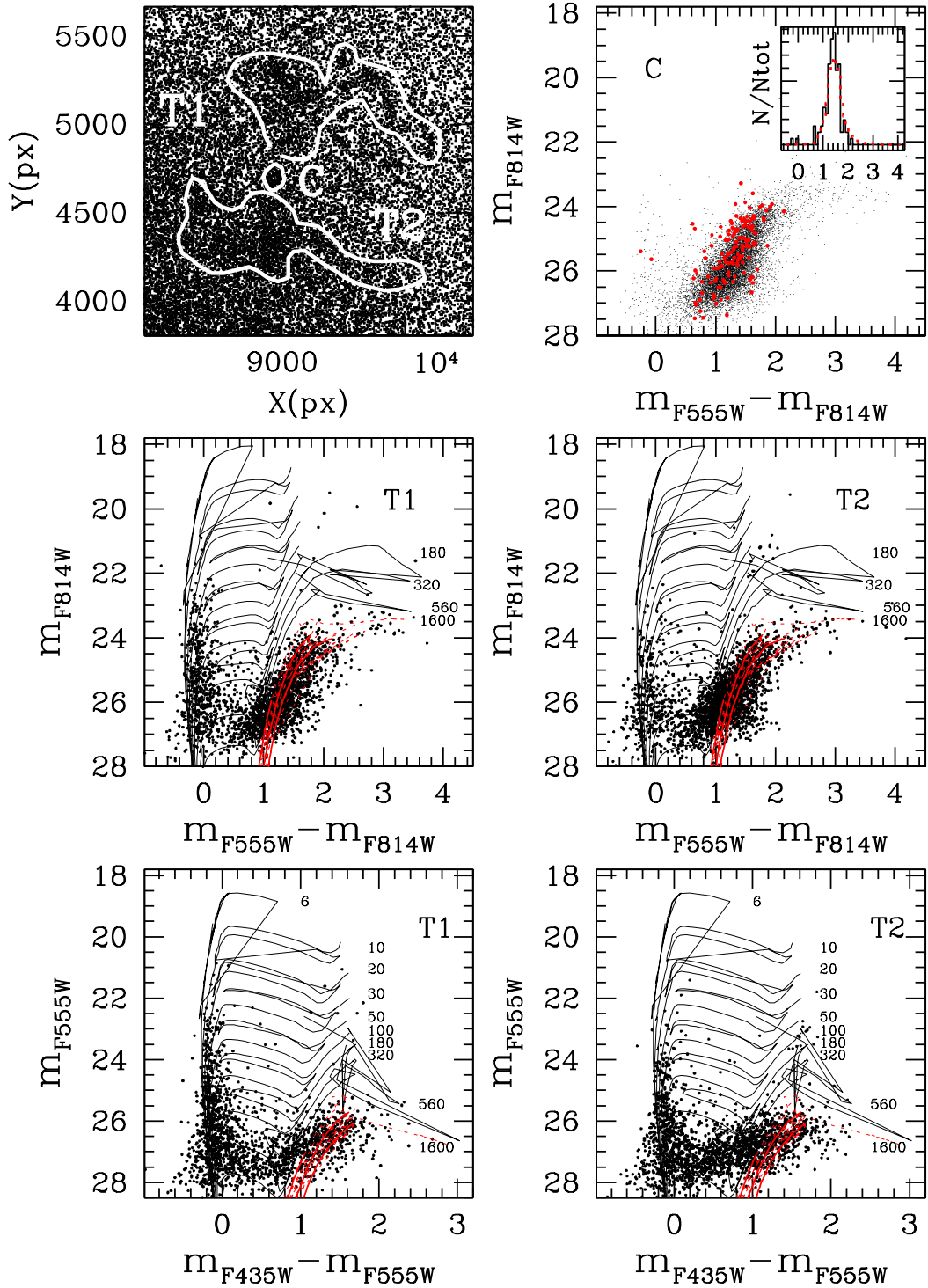}
\caption{CMDs for the stars resolved in the outskirts of CL~77 and in  two regions (T1 and T2) outlined to include the blue stars in the tails.
Top left: spatial distribution of the stars in a (60$\times$60) arcsec$^2$  
field containing regions C, T1, and T2. Top right: CMD of CL~77 (big red points) superimposed 
to the CMD of the stars resolved in region (5;4) of A08 (small dots). An insertion shows the color distributions of the cluster and field stars down to 
2 magnitudes below the RGB tip. Middle and bottom panels: CMDs of T1 and T2, with superimposed 
the Padova \citep{pad10} isochrones for log(Age)=6.75, 7.00, 7.25....7.75, 9.20, 9.50, and 10.00. Isochrones older than 1 Gyr are plotted  
using the red solid line up to the RGB tip phase, and the red dashed line for later phases.  
Labels indicate ages in Myr.
 \label{sat_cmd}}
\end{figure*}
\end{center}

Fig.~\ref{sat_img} shows a color-combined image centered on CL~77 and 
including the blue tails. The cluster is partially resolved into stars at its outskirts in the I image, but it is  only poorly resolved in the B image.
The CMDs of the stars resolved in an elliptical region (C) around CL~77 and in two regions including the tails (N-W tail: T1;  S-E tail: T2) are shown in Fig.~\ref{sat_cmd}.   The T1 and T2 contours were outlined following the distribution of the blue/young stars in Fig.~1 and in the bottom left panel of Fig.~3.
The cluster stars are almost all RGB stars (top right panel of Fig.~\ref{sat_cmd}).
A comparison with a nearby field 
indicates no significant difference between the properties of the two RGBs. 
The CMD confirms that CL~77 is older than $\sim$ 1 Gyr (as inferred from the integrated colors in A11), and shows that the RGB color is consistent with a metallicity of Z$\sim$0.004, as for the field RGB stars (see A08 and CMDs in the middle and lower panels of Fig.~\ref{sat_cmd}). 

The CMDs of T1 and T2 are shown in the middle and lower panels of Fig.~\ref{sat_cmd}, respectively, with the Z$=$0.004 Padova isochrones \citep{pad10} superimposed. 
Both diagrams display old (age$>$1 Gyr) RGB stars, as well as populated ``blue plumes'' at  
V$-$I, B$-$V $\sim-0.1$ 
and red supergiants, indicative of recent star formation (SF). In T1, the blue plume is well populated up to $V\sim 22.5$, and there are a few blue supergiants as bright as 
$V\sim21$; the red supergiants at $1.2\lesssim B-V \lesssim1.8$ and $21\lesssim V \lesssim22.5$ are consistent 
with ages between 50 and 20 Myr ago. 
The blue loop and AGB phases for ages between  $\sim$50 Myr and 1 Gyr appear scarcely populated, 
suggesting that the SF was not very active in T1 at ages older than $\sim$50 Myr. 
In T2, the bulk of the recent SF seems to be older than in T1, because the drop in the blue luminosity function occurs at a fainter magnitude ($V\sim24$) than in T1. 
In T2, we also observe a concentration of AGB stars at  $1.4\lesssim B-V \lesssim1.8$ and $22.5\lesssim V \lesssim23.5$, corresponding to isochrones of ages between 
$\sim$200 and $\sim$50 Myr. This clearly identifies a burst of SF occurred in T2 between $\sim$200 and $\sim$50 Myr ago. The SF possibly continued at a lower level until $\sim$20 Myr ago.  The CMDs indicate a lower level of SF between $\sim$200 Myr and $\sim$1 Gyr ago.  

We attribute the old ($>$1 Gyr) RGB stars in T1 and T2 to the NGC~4449 field. In fact, as shown in the bottom right panel of 
Fig.~\ref{wave}, there is no evidence for an overdensity of old stars in correspondence of the tails. 

\section{Data at other wavelengths}

\begin{center}
\begin{figure}
\epsscale{1.2}
\plotone{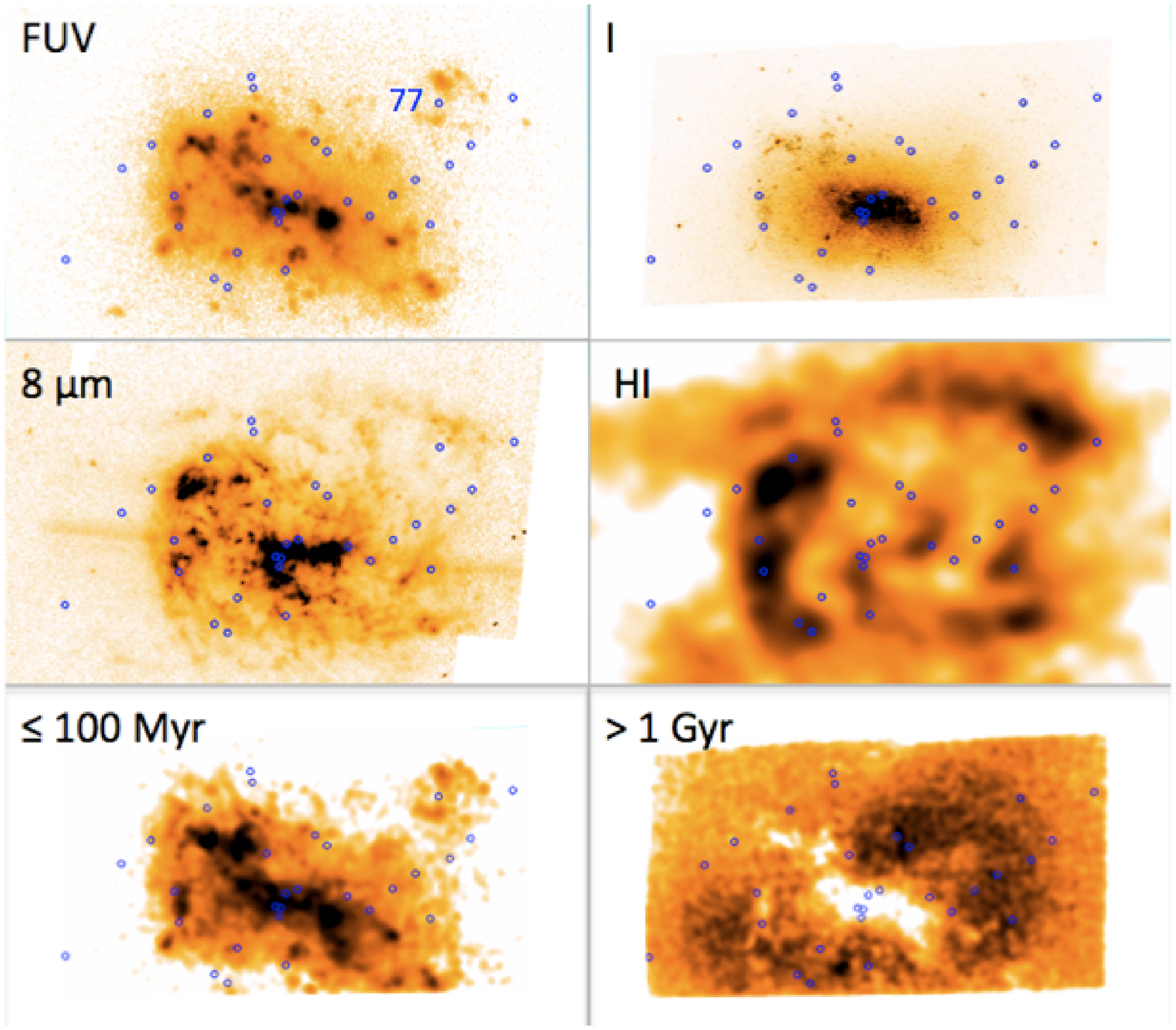}
\caption{On the same scale, images of NGC~4449 at different wavelengths: FUV-GALEX (top left), I-ACS (top right), 
8 $\mu$m-Spitzer (middle left), HI-VLA (middle right). 
In the bottom panels are density maps of the stars younger than 100 Myr 
(left) and older than 1 Gyr (right). 
Overplotted are the clusters older than 1 Gyr identified by A11. CL~77 is surrounded by two tails well visible in the FUV image and in the young star 
density map. 
 \label{wave}}
\end{figure}
\end{center}

Figure~\ref{wave} shows, on the same scale of our I ACS mosaicked image, the FUV (GALEX),  8.0$\mu$m ({\it Spitzer}), and HI 
(VLA) images. The FUV data are from the 11HUGS survey  \citep[][]{hugs} 
and trace stars younger than $\sim$100 Myr. The 8.0$\mu$m image, which traces polycyclic aromatic hydrocarbons (PAHs),
belongs to the {\it Spitzer} LVL survey \citep[][]{lvl}.
The HI integrated intensity map is from the THINGS survey  
\citep[][]{things}. We also show in Fig.~\ref{wave} density maps of the resolved stars younger than 100 Myr and older than 1 Gyr derived from the CMDs in A08. 

\begin{center}
\begin{figure}
\epsscale{1.}
\plotone{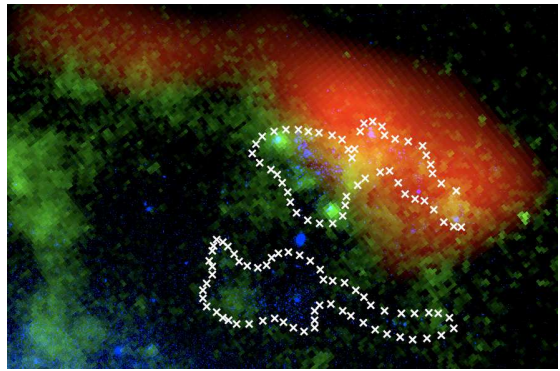}
\caption{B (blue), 8$\mu$m (green), and HI (red) color-composite image of a (130$\times$90) arcsec$^2$ field including CL~77 and the two tails. 
The same contours used to define T1 and T2 in Fig.~2a are overlaid. 
 \label{birachi}}
\end{figure}
\end{center}

In the FUV, the tails around CL~77  
appear as two diametrically opposed, symmetric structures on which the cluster is centered. 
However, the 8$\mu$m and HI images show that they are in fact not
symmetric: while T2 does not appear very prominent in these bands, T1 is associated with significant emission at 8$\mu$m
(as well as at longer infrared wavelengths, not showed here), with one of the brightest HI spots in the whole galaxy, and also with ionized gas \citep[see Fig.~1 in][]{hunter05}. Fig.~\ref{birachi} 
shows that the peak of the HI is offset with respect to the stars in T1 and to the infrared emission (and also with respect to the ionized gas). From the integrated HI map, and using equations (1) and (5) in \cite{things}, we estimate an average column density as high as 
$N_{HI}\sim3.4\times10^{21} cm^{-2}$ within a circular aperture of $\sim$10 arcsec radius 
centered on the HI peak. 
It is also interesting that the infrared and HI emissions associated with T1 appear as part of a much longer stream 
wrapping around the galaxy from west to north. The HI stream 
appears to follow the rotation of the more external HI disk which 
counter-rotates with respect to the internal HI disk  \citep[Fig.~14 in][]{hunter99}. 


\section{Summary and Discussion \label{discussion}}


 A11 showed that CL~77 is $\sim$ 7 Gyr old, massive ($M\sim1.7\times10^6 M_{\odot}$), compact 
($r_e\sim1.9$ pc) and flattened ($\epsilon\sim0.24$). Two other GCs in  NGC~4449 are as massive and elliptical as CL~77 (CL~34 and CL~79 in A11), with fairly similar CMDs and color distribution of their RGB stars. CL~77
is  however peculiar because it is located at the projected center of a symmetric structure outlined by blue stars reminiscent of tidal tails. CL~77
 is the only GC in NGC~4449 that appears associated to such a structure. The presence of blue tails is unusual also because the cluster is located far away from the galaxy center ($\sim$ 3 kpc) and  from any region of active SF.  Moreover, in a diagram of mean projected mass density inside $r_e$ versus total mass  \citep[Fig.~3 in][]{walcher05}, it lies at the upper end of the GC distribution, where galaxy nuclei are.
 Hence, a possible interpretation is that CL~77 is what remains of a dwarf galaxy being disrupted by NGC~4449, similarly to what is suggested for $\omega$Cen and G1 \citep[e.g.][]{bekki03,meylan01}.

The stars resolved in the outskirts of the cluster confirm that it is old and indicate that its metallicity is consistent with 
$\sim$1/4 solar, i.e. similar to that of NGC~4449. The blue tails host recent SF:  T2, closer to the galaxy body, exhibits a SF episode between $\sim$200 and $\sim$ 50 Myr ago; T1 hosts younger SF than in T2, occurred until $\sim$20 Myr ago, and possibly still active as indicated by the presence of ionized gas \citep{hunter90}. 
T1 is also associated with PAH and dust emission, and with one of the brightest HI spots in the galaxy ($N_{HI}\sim3.4\times10^{21} cm^{-2}$). This emission appears as part of a much longer stream which wraps around the galaxy, and which counter-rotates with respect to the inner HI disk.


The presence of recent SF in the tails suggests that supposed accreted galaxy had plenty of gas; 
thus it probably was a dwarf irregular, a small late-type spiral, or an HI-rich dwarf elliptical (dE), like, e.g., those discovered in the Virgo cluster by \cite{conselice03}.
Furthermore, T1 is associated with one of the brightest HI spot in NGC~4449, which may be gas belonging to the accreted galaxy.
Another possibility is that this gas was already present in NGC~4449; however its high density suggests that compression may have occurred, and this 
requires interaction with another gas-rich system.

Rossa et al. (2006) showed that, for spiral galaxies, the mass of the central nuclear star cluster correlates with the mass/luminosity of the bulge, and the same result was found by \cite{wehner06} for dE galaxies. 
The correlation is such that $\log M_{cluster}= (1.04\pm0.06)\times \log M_{bulge}-(3.10\pm0.60)$. 
From $M_{CL77}\sim1.7\times10^6 M_{\odot}$ we derive that the parent galaxy bulge or spheroid had a mass of 
 $\sim0.93^{+3.4}_{-0.73} \times10^9 M_{\odot}$. This is the total mass, from both luminous and dark matter, contained within the central spheroidal component of the galaxy, and it is largely dominated by the luminous mass. If the galaxy was a spiral with substantial disk, then the total galaxy mass may have been even larger. 
Comparing this value with NGC~4449's total stellar mass of $\sim 3\times10^9 M_{\odot}$, as derived 
from the SF history in \cite{mcq10}, we obtain a mass ratio of $\sim$ 3:1  (but notice that  the large scatter in the \cite{wehner06} relation implies 
mass ratios in the range 15:1 to 0.7:1, considering the 1$\sigma$ error in the satellite mass).
For comparison, typical (visible) masses often quoted for the LMC and SMC are 
$\sim2\times10^{10} M_{\odot}$ and $\sim2\times10^{9} M_{\odot}$, respectively \citep[see e.g.,][Section~5.1]{besla07} resulting in a LMC:SMC ratio of 10:1.  


Admittedly, from the available data, we cannot prove that  T1 and T2 are physically associated with CL~77, and
the possibility  that we are seeing the chance superposition along the line of sight of SF regions with a cluster associated to the old spheroid of NGC~4449, although unlikely, can not be ruled out. In the following we discuss some of the possible questions.

{\it Why is there no evidence in our images for streams or tails of old stars associated with CL~77?} 
Indeed, this appears as a problem for our proposed scenario, since if the progenitor 
had a stellar mass of $\sim10^{9} M_{\odot}$, the satellite remnant should have a surface brightness 
 $\mu_{I} \lesssim$22 mag/arcsec$^2$ (obtained in the assumption that the satellite stars are uniformly distributed over our ACS field of view). 
Since in the proximity of CL~77, $\sim$ 3 kpc away from the galaxy center, 
we measure $\mu_{I}\sim$23 mag/arcsec$^2$
(but significantly brighter in the regions of the blue tails  due to the contribution from young luminous stars), 
we should be able to detect streams associated to the disrupted galaxy. 
However, this problem may be solved considering the large scatter in the \cite{wehner06} relation and the factor $\sim$5 uncertainty in the progenitor mass. 
For instance, if the satellite mass was as low as $2\sim10^{8} M_{\odot}$, its remnants would be probably too faint to be detected against the NGC~4449   background. 
Interestingly, a stellar stream was recently discovered in NGC~4449 \citep[][in prep.]{delgado11}, but that stream is located on the opposite side of the galaxy from CL~77.

The blue tails extend $\sim$ 50'' in width, corresponding to $\sim$ 1 kpc at NGC~4449's distance.
{\it Could the stripped debris spread so rapidly to occupy such a large width if it were originally more compact?}
N-body simulations show that, just after 
the satellite orbital pericenter, tidal tails are highly diffuse \citep[see e.g. Fig.~4 in][]{klim09}.
Given the small (projected) distance of CL~77 from NGC~4449's center, this possibility can not be ruled out. 
Another possibility is that the accreted satellite was a very-low surface brightness disk galaxy
(like e.g. Maffei I), with a lot of gas.  The blue tails could be the result of star formation resulting from gas compression in the
interaction (not ``traditional'' tidal tails in the sense of material stripped off the tidal radius of the progenitor).

{\it How might the outer regions of CL~77's progenitor be completely stripped off over a long interaction history but fail
  to strip the gaseous component until recently?}
 A possible answer comes from simulations of gas-rich dwarf galaxy satellites orbiting within a MW-sized halo accounting for the combined effects of tides 
  and ram pressure stripping caused by a hot diffuse gaseous corona surrounding the MW \citep[e.g.][]{mayer06}. They show
   that dwarfs with dark haloes more massive than 
  $\sim 10^9 M_{\odot}$ lose most  of their gas  only if a heating source (e.g. the cosmic ultraviolet background at $z>2$)
  keeps the gas extended. Galaxies falling into the MW halo at lower redshift can retain significant amounts of the centrally concentrated gas.

 
To summarize, we consider the following properties as indications that CL~77  may be  the nucleus of a former gas-rich satellite galaxy undergoing disruption by NGC~4449:
 a) the cluster is associated with two tidal tails of young (age$\lesssim$200 Myr) stars;  b) in a diagram of mean projected mass density inside $r_e$ versus total mass
it occupies the locus of galaxy nuclei; c) one of the tail 
is associated with a concentration of HI and infrared (dust/PAHs) emission which appears as part of a much longer stream wrapping around the galaxy. In addition, CL~77 is highly massive and elliptical. Its  mass is not too far from the values derived in the literature for $\omega$Cen 
\citep[2.5 $\pm$ 0.3 $\times 10^6 M_{\odot}$,][]{vandeven06}, 
and its ellipticity is comparable to those of $\omega$Cen and G1 \citep[$\epsilon\sim0.2$,][]{pancino00,meylan01}. 
 Kinematical data (for CL~77, for other clusters in NGC~4449, and possibly for some bright blue stars in the tails) 
or deeper imaging would be useful to further test the scenario in which CL~77 is the nucleus of an accreted galaxy.

CL~77 is of interest in that it may show the process of satellite
disruption in a different phase than other known GCs suspected to be
former galactic nuclei. 
It has been suggested \citep{bellazzini08,carretta10} that $\omega$Cen and M54 might have followed a similar evolutionary path, with M 54 seen in an earlier phase compared to $\omega$Cen. 
CL~77 may represent an intermediate stage between M54, where the galaxy stream is still visible, and $\omega$Cen or G1, where the presumed disrupted galaxy is no longer visible.  By putting the various clusters in sequence we
can create a times series of events composed of single-epoch
``snapshots''. M54 is still surrounded by a parent galaxy that can be
recognized as such (the Sgr dSph), but which is being drawn out into
extended streams. CL~77 does not show a surrounding parent galaxy
anymore, but it has surrounding features that hint at its former
presence (the blue tails) as well as association with an extended (HI
and infrared) stream. Palomar 14 shows neither a surrounding parent
galaxy nor an extended stream anymore, but short tails in the
immediate vicinity of the cluster's tidal radius are still
evident. And finally, $\omega$Cen and G1 show neither a surrounding parent
galaxy, nor extended streams, nor local tails. In $\omega$Cen, only the
presence of multiple stellar populations may still point to its former
location at a galaxy center.

\acknowledgments

We thank R. Sancisi, F. Fraternali, M. Bellazzini and M. Mapelli for useful discussions.  
We thank the anonymous referee that helped to improve the presentation and balance of the paper. 
Support for proposal \#10585 was provided by NASA
through a grant from STScI which is operated by
AURA    Inc.    under NASA contract NAS 5-26555.
FA and MT have received partial financial support from ASI, through contract COFIS
ASI-INAF I/016/07/0 and contract ASI-INAF I/009/10/0.
This work made use of THINGS, The HI Nearby Galaxy Survey \citep{things}.






\appendix





\clearpage




\end{document}